\documentstyle[12pt]{article}

\newcommand{\be}{\begin{equation}}
\newcommand{\ee}{\end{equation}}
\newcommand{\bea}{\begin{eqnarray}}
\newcommand{\eea}{\end{eqnarray}}

\renewcommand{\theequation}{\arabic{section}.\arabic{equation}}

\begin{document}
\begin{titlepage}

\begin{flushright}
hep-th/9705050\\
SUSX-TH-97-007\\
PU-RCG 97/5
\end{flushright}

\vspace{.4in}

\begin{center}
\Large
{\bf  S-duality invariant perturbations in string cosmology}

\vspace{.4in}

\normalsize
 
\large{E.~J.~Copeland$^{1a}$ , James E.~Lidsey$^{2b}$ and David Wands$^{3c}$}

\normalsize
\vspace{.3in}

$^1${\em Centre for Theoretical Physics, University of Sussex, Brighton, 
BN1 9QH, U. K.} 

\vspace{.2in}

$^2${\em Astronomy Unit, School of Mathematical 
Sciences,  \\ 
Queen Mary \& Westfield, Mile End Road, LONDON, E1 4NS, U. K.}

\vspace{.2in}

$^3${\em School of Mathematical Studies, University of Portsmouth, \\ 
Portsmouth, PO1 2EG, U. K.}

\end{center}

\vspace{0.4in}

\baselineskip=24pt
\begin{abstract}
\noindent
We investigate the generation of curvature and isocurvature (dilaton,
moduli and axion) perturbations in a general class of
axion--dilaton--moduli models, including the pre--big bang
scenario. Allowing for an arbitrary coupling constant $\lambda$ between
the dilaton field and the axion field, we exploit the SL(2,R)
symmetry of the theory to obtain the spectral indices of the field
perturbations in a pre--big bang type scenario. Axion field
fluctuations about a homogeneous background field can yield a
scale-invariant (Harrison-Zel'dovich) spectrum. As an example we
present a string-motivated case with SL(2,R)$\times$SL(2,R) symmetry, where a
second axion field arises from the compactification of the 
ten--dimensional theory to four dimensions.

\end{abstract}


\vspace{.4in}
$^a$Electronic address: e.j.copeland@sussex.ac.uk 

$^b$Electronic address: jel@maths.qmw.ac.uk

$^c$Electronic address: wands@sms.port.ac.uk

\end{titlepage}


\section{Introduction}

\setcounter{equation}{0}

\def\theequation{\thesection.\arabic{equation}}

The cosmology of string theory is presently a very active area of
research.  The early universe provides one of the few areas where we
may be able to place observational constraints on string theory. In
particular one would like to know whether an inflationary expansion
could occur in string theory which is capable of producing the initial
conditions required by the standard hot big bang model.  Recently, an
alternative inflationary model, known as the {\em pre--big bang
scenario}, has been developed based on the low--energy string
effective action~\cite{pbb}.

The fundamental assumption in the pre--big bang scenario is that the
universe is initially in a weakly--coupled, low curvature regime. The
non--minimal coupling between the dilaton and graviton can lead to 
a super--inflationary expansion, where 
the Hubble radius decreases with time
and the universe evolves into a state of
high curvature and strong coupling. This scenario has a number of
interesting features compared with conventional inflation
models. Perturbations arising from quantum fluctuations can be
calculated exactly in many models without resorting to `slow--roll'
approximations and their amplitude may be normalised in the low
curvature, weakly--coupled regime, where the vacuum state is well
understood \cite{gasperini,em,CEW97}.  Moreover, there is 
no need to introduce a potential energy term during
inflation~\footnote{For a more sceptical view of the pre--big bang
scenario see~\cite{TW97}}.

However, there are a number of pressing problems in this scenario
which need to be addressed. Perhaps the most fundamental is the
mechanism by which the singularity present in the high curvature
regime can be avoided, bringing the superluminal expansion to an end
(the Graceful Exit Problem)~\cite{gracefulexit}. For recent work on
possible resolutions see~\cite{lidsey,solutions}. Another important
issue is whether density perturbations can be generated that are
capable of seeding the observed large--scale structure in the
universe. It is known that the dilaton and moduli fields produce
perturbations with spectral index $n=4$, making them too steep to be
viable seeds for large-scale structure~\cite{gasperini}. One
possibility is that the electromagnetic perturbation spectrum
generated in the pre--big bang could have interesting observational
consequences~\cite{em}.

Recently, Copeland, Easther and Wands \cite{CEW97} have shown that the
coupling of the dilaton to the Neveu--Schwarz/Neveu--Schwarz (NS--NS)
axion field in four dimensions can produce an axion spectrum which
includes the scale--invariant $n=1$ case. The precise value depended
on the expansion rate of the internal dimensions and varied between
$0.5<n \leq 4$.  In that work, the role of the SL(2,R) 
``S--duality'' symmetry of the
action was crucial. It was possible to construct the unique S-duality
invariant field perturbations for the axion and dilaton fields and
from them derive S-duality invariant solutions to the perturbation
equations, valid when the axion field is time-dependent as well as in
the case when it was frozen.  In this paper we extend these ideas,
considering the generation of perturbations about homogeneous
background solutions in a more general class of models that may be
relevant to the pre--big bang scenario. Once again we will find
S-duality plays a significant role in enabling us to obtain
perturbation spectra for the fields.
 
We consider an effective Jordan--Brans--Dicke--type theory of gravity
\cite{BransDicke61}:
\be
\label{actionone}
S=\int
d^4 x \sqrt{-g}e^{-\Phi} \left[ R -\omega \left( \nabla \Phi 
\right)^2 -\frac{1}{2} e^{\Gamma \Phi}  \left( \nabla \sigma \right)^2 
- \frac{1}{2} \left( \nabla \beta \right)^2  \right] \ ,
\ee
where $R$ is the Ricci scalar of the space--time with metric
$g_{\mu\nu}$ and $g = {\rm det} g_{\mu\nu}$. Our metric has signature
$(-,+,+,+)$.  $\Phi$ represents the dilaton field, $\beta$ is a
`modulus' field and $\sigma$ is an `axion' field,
distinguished from the moduli by its explicit coupling to the dilaton.
The axion-dilaton coupling is determined by the constant $\Gamma \ne 0$.
The Brans-Dicke constant $\omega$ determines the dilaton--graviton
coupling.
The region of parameter space where super--inflationary expansion 
in a pre--big bang phase may 
occur is $-4/3 < \omega  < 0$ \cite{lidsey}. 

An effective action such as that given in Eq.~(\ref{actionone}) can
arise in various theories. One may consider the toroidal
compactification of higher--dimensional Einstein gravity, i.e.,
Kaluza-Klein theory. If the gauge fields remain frozen and the
internal space is isotropic with radius $b$ and dimensionality $d$,
the resulting four--dimensional action is given by the
dilaton--graviton sector of Eq. (\ref{actionone}) with $\Phi \propto
-d \ln b$ and $\omega =-1 +1/d$.  In the string effective action the
dilaton already appears in the higher-dimensional theory and
compactification to four-dimensions leads to a dilaton--graviton
sector with $\omega =-1$ plus extra moduli fields represented by the
field $\beta$ \cite{effectiveaction}.  When $\Gamma =1$, $\sigma$ may
be viewed as a massless scalar field in the matter sector of the
theory~\cite{MW95}. When $\Gamma = 2$ the gradient of the
pseudo-scalar axion field is dual to the field strength of the
two--form potential arising in the NS--NS sector of the string
effective action~\cite{Behrndt,CLW94}. Other couplings for an
axion-type field are possible when moduli arising from the
compactification of form fields on the internal space are
excited~\cite{sc1}.

This paper is organised as follows. In Section 2, we discuss the
global SL(2,R) symmetry of action Eq.~(\ref{actionone}) and employ
this symmetry to derive the general spatially flat
Friedmann--Robertson--Walker (FRW) cosmology in Section 3. In Section
4, we derive the perturbation spectra of the three scalar fields that may
arise from quantum fluctuations about this homogeneous background. In
Section 5 we provide a string motivated example of the spectra
obtained including moduli fields arising from compactification of the
low--energy string action from ten to four dimensions on a
six-dimensional torus. This possesses an additional SL(2,R) symmetry
\cite{witten,gs}. We conclude in Section 6.

\section{Global Symmetry of the Effective Action}

\setcounter{equation}{0}

\def\theequation{\thesection.\arabic{equation}}

The global symmetry of theory Eq.~(\ref{actionone}) becomes apparent in the
`Einstein' frame where  the dilaton field is minimally coupled to
gravity.  This is obtained by the conformal
transformation
\be
\label{conformal}
\tilde{g}_{\mu\nu} =\Omega^2 g_{\mu\nu}, \qquad \Omega^2 \equiv 
e^{-\Phi}.
\ee
We will refer to the original frame with 
metric $g_{\mu\nu}$ as the Jordan frame.\footnote{For the examples 
which refer directly to string theory this will correspond to 
the string frame.}
If we also redefine the dilaton field:
\be 
\label{fieldredefinition}
\phi \equiv (3+2\omega)^{1/2} \Phi \ ,
\ee
the action Eq.~(\ref{actionone}) transforms to
\be
\label{einsteinactionone}
S=\int d^4x \sqrt{-\tilde{g}} 
\left[ \tilde{R} -\frac{1}{2} \left( \tilde{\nabla} \phi \right)^2 
-\frac{1}{2} e^{2\lambda \phi} \left( \tilde{\nabla} \sigma \right)^2 
-\frac{1}{2} \left( \tilde{\nabla} \beta \right)^2 \right]   ,
\ee
where the axion--dilaton coupling in the Einstein frame is
\be
\label{lambda}
\lambda  \equiv  \frac{\Gamma}{2(3+2\omega )^{1/2}} .
\ee

Defining the symmetric $2\times 2$ matrix
\be
\label{S}
N \equiv   \left( \begin{array}{cc} e^{\lambda \phi} &  \lambda \sigma 
e^{\lambda \phi} \\ 
\lambda \sigma e^{\lambda \phi} & e^{-\lambda \phi} +\lambda^2
\sigma^2 e^{\lambda \phi} \end{array} \right)
\ee
then implies that action Eq.~(\ref{einsteinactionone}) may 
be written as
\be
S=\int d^4x \sqrt{-\tilde{g}} \left[ \tilde{R} +\frac{1}{4 \lambda^2}
{\rm Tr} \left( \tilde{\nabla} N \tilde{\nabla} N^{-1} \right)
-\frac{1}{2} \left( \tilde{\nabla} \beta \right)^2 \right] \ .
\ee
The matrix $N$ is a member of the group SL(2,R), and thus
$N^{-1}=-JN^TJ$,  where
\be 
J = \left( \begin{array}{cc} 0 & 1 \\ -1 & 0 \end{array} \right)
\ee
is the SL(2,R) metric. The dilaton and axion fields parameterise an
SL(2,R)/U(1) coset and the action is invariant under a global 
SL(2, R) symmetry:
\be
\tilde{g}_{\mu\nu} \to \tilde{g}_{\mu\nu}\ , \quad
N\to \Theta N \Theta^T \ , \quad
\beta \to \beta \ ,
\ee
where
\be
\Theta = 
\left( \begin{array}{cc} d  & c \\ b & a 
\end{array} \right)
\ee
is a member of SL(2,R) and hence $ad-bc=1$.

The SL(2,R) group acts non--linearly on the complex scalar field $\chi
= \lambda \sigma + ie^{-\lambda  \phi}$ such that the transformed 
field $\chi \to (a \chi
+b)/(c \chi +d)$. Hence the transformed scalar fields are given by
\bea
\label{nonlin1}
e^{\lambda \phi} 
\to c^2 e^{-\lambda \phi} + \left( d + c \lambda 
\sigma \right)^2 e^{\lambda \phi} \\
\label{nonlin2}
\lambda \sigma e^{\lambda \phi} \to 
ac e^{-\lambda \phi} + 
e^{\lambda \phi} (b + a \lambda \sigma )( 
d + c \lambda \sigma ) \ . 
\eea
When $\lambda \sigma= -d/c$ and $c^2=1$ this reduces to
$\phi\to-\phi$ and is a strong-weak coupling ``S-duality''. 
The modulus field remains invariant under the SL(2, R) transformation. 
The Einstein frame Lagrange density including the modulus field is also
invariant under changes in the sign of the modulus field,
$\beta\to-\beta$, or a global shift $\beta\to\beta+$constant. 

Since Eqs. (\ref{conformal}) and (\ref{fieldredefinition}) correspond
simply to field redefinitions, the SL(2, R) symmetry of the theory  is
also maintained in the Jordan frame.    
However the metric in the Jordan frame does change due to
the change in the conformal factor:
\be
\Omega^2 \equiv e^{-\Phi} \equiv e^{-\phi/\sqrt{3+2\omega}} \to 
 \left[ c^2 e^{-\lambda\phi} + \left( d + c \lambda\sigma \right)^2
e^{\lambda\phi} \right]^{-1/(\lambda\sqrt{3+2\omega})} \ .
\ee

\section{Homogeneous solutions}

\setcounter{equation}{0}

\def\theequation{\thesection.\arabic{equation}}

In this Section we present the general solutions for a
spatially flat FRW cosmology~\cite{jim}, generalising earlier results
found for specific couplings between the
fields~\cite{CLW94,Mukherji}. These homogeneous axion-dilaton
solutions can readily be extended to non-flat FRW models~\cite{CLW94}
and anisotropic models~\cite{MW95,Batakis,jdb96}.

We will work in the Einstein frame where the line element is given by
\be
d\tilde{s}^2 = \tilde{a}^2(\eta) \left( -d\eta^2 + dx^2 + dy^2 + dz^2
\right) \ ,
\ee
where $\eta$ is the conformally invariant time coordinate.
The equations of motion are 
\begin{eqnarray}
\phi'' + 2h\phi' &=& \lambda e^{2\lambda\phi}\sigma'^2 \ ,\\
\sigma'' + 2h\sigma' &=& -2\lambda\phi'\sigma' \ ,\\
\beta'' + 2h \beta' &=& 0 \ ,\\
h' + 2h^2 &=& 0 \ ,
\end{eqnarray}
where $h\equiv \tilde{a}'/\tilde{a}$ and a prime denotes differentiation
with respect to conformal time. We also have the Friedmann constraint
equation
\begin{equation}
h^2 = {1\over12} \left( \phi'^2 + \beta'^2 + e^{2\lambda\phi}\sigma'^2
\right) \ .
\end{equation}

It is straightforward to derive the general dilaton-moduli-vacuum
solutions where $\sigma'=0$. All the remaining equations of
motion can be integrated directly and one obtains
\begin{eqnarray}
\label{vacuumsol}
\tilde{a} &=& \tilde{a}_* \left| {\eta\over\eta_*} \right|^{1/2} \ ,\\
\label{vacuumphisol}
e^\phi &=& e^{\phi_*} \left| {\eta\over\eta_*} \right|^{r_\pm} \ ,\\
\label{vacuumbetasol}
e^\beta &=& e^{\beta_*} \left| {\eta\over\eta_*} \right|^s \ .
\end{eqnarray}
The Friedmann constraint equation requires $r_\pm=\pm\sqrt{3-s^2}$. 

These dilaton-moduli vacuum solutions are monotonic power-law
solutions.  The solutions with $r_+>0$ approach weak coupling
($\phi\to-\infty$) as $\eta\to0$ and strong coupling
($\phi\to+\infty$) as $\eta\to\pm\infty$, while the reverse happens on
the $r_-<0$ solutions. The $r_+$ and $r_-$ solutions are related by
the S-duality $\phi\to-\phi$.  However, for either $r_+$ or $r_-$, the
solutions in the Einstein frame always evolve from a low curvature
regime as $\eta\to-\infty$ towards a high curvature regime as
$\eta\to0$, or from a high curvature regime as $\eta\to0$ to a low
curvature regime as $\eta\to+\infty$.

The general SL(2,R) transformation given in Eqs.~(\ref{nonlin1})
and~(\ref{nonlin2}) then allows one to write down the general solution
for a spatially flat FRW universe with
$\sigma'\neq0$~\cite{jim,CLW94,Mukherji}:
\begin{eqnarray}
\label{axia}
\tilde{a} &=& \tilde{a}_* \left| {\eta\over\eta_*} \right|^{1/2} \ ,\\
\label{axiphi}
e^\phi &=&  {e^{\phi_*} \over 2^{1/\lambda}}
 \left[ \left| {\eta\over\eta_*} \right|^{-\lambda r}
   + \left| {\eta\over\eta_*} \right|^{\lambda r} \right]^{1/\lambda} \ ,\\
\label{axisig}
\sigma &=& \sigma_* \pm {e^{-\lambda\phi_*} \over \lambda}
 \left[ { |\eta/\eta_*|^{-\lambda r} - |\eta/\eta_*|^{\lambda r}
  \over |\eta/\eta_*|^{-\lambda r} + |\eta/\eta_*|^{\lambda r} } \right]
\ ,\\
\label{axibeta}
e^\beta &=& e^{\beta_*} \left| {\eta\over\eta_*} \right|^s \ ,
\end{eqnarray}
where we still have $r^2+s^2=3$, although the sign of $r$ is now
irrelevant and we shall choose $r>0$ in what follows. 

The evolution of the axion and dilaton in the Einstein frame is
determined solely by the constant $\lambda$. The asymptotic form
of the solutions in the early or late times limits (either
$\eta\ll\eta_*$ or $\eta\gg\eta_*$) is determined by the
dilaton--moduli--vacuum solutions given in
Eqs.~(\ref{vacuumsol})--(\ref{vacuumbetasol}),  where $\sigma'\to0$. 

For $\lambda>0$ the general axion-dilaton-moduli solution approaches
the dilaton--moduli--vacuum solution with $r_-=-r$ when
$\eta\to0$. The behaviour for $|\eta|\gg|\eta_*|$ is given by the
(duality related) vacuum solution with $r_+=+r$. The effective
gravitational coupling $G_{\rm eff} \propto
e^{\Phi}=e^{\phi/\sqrt{3+2\omega}}$ is bounded from below. It is
divergent as $\eta\to0$ and decreases to a minimum value at a time
$\eta=\eta_*$. It then increases indefinitely as
$|\eta|\to\infty$. The lower bound on the coupling is determined by
the canonical momentum of the axion field.

The qualitative behaviour of the cosmology is similar to that derived
originally for the string model when the axion is dual to the NS--NS
three--form field strength and $\lambda =1$~\cite{CLW94}.  Our
analysis shows that the qualitative behaviour does not depend too
sensitively on the coupling between the dilaton and axion fields and
is therefore quite generic in theories of this type. However the
asymptotic limits are interchanged for $\lambda<0$. The $r_+=+r$
solution applies for $|\eta| \ll |\eta_*|$ and the $r_-=-r$ branch at
$|\eta| \gg |\eta_*|$.  The coupling $\phi$ is now bounded from
above. The effective gravitational coupling $e^\Phi$ increases from
zero at $\eta\to0$, reaches a maximum at $\eta=\eta_*$, and decreases
back to zero as $|\eta|\to\infty$.

The value of $\sqrt{3+2\omega}$ is important in determining the
behaviour of the scale factor in the Jordan frame.  The conformal
transformation presented in Eq.~(\ref{conformal}) leads to a relation
between the cosmological scale factor, $a$, in the Jordan frame and the
scale factor in the Einstein frame, 
\be
\label{einsteina}
a = e^{\phi/(2\sqrt{3+2\omega})} \tilde{a} \ .
\ee
Using Eqs.~(\ref{vacuumsol}) and~(\ref{vacuumphisol}) then implies that
when $\sigma'=0$ the scale factor in the Jordan frame is
\begin{equation}
\label{vacuumasol}
a = \tilde{a}_*e^{\phi_*/(2\sqrt{3+2\omega})} \left| {\eta\over\eta_*}
\right|^{(1+r_\pm/\sqrt{3+2\omega})/2} \ .
\end{equation}
For $\sigma'\neq0$ the Jordan frame scale factor evolves as 
\be
\label{gensol}
a =a_* \left| \frac{\eta}{\eta_*} \right|^{1/2} 
\left[ \left| \frac{\eta}{\eta_*} \right|^{\lambda r} + 
\left| \frac{\eta}{\eta_*} \right|^{- \lambda r}
\right]^{1/(2\lambda\sqrt{3+2\omega})}  \ .
\ee

For $\lambda>0$ and $r>\sqrt{3+2\omega}$ the scale factor in the
Jordan frame is infinitely large as $\eta\to0$ and the Universe
undergoes an accelerated contraction for $\eta>0$. It reaches a
minimum size before re-expanding to infinity as
$\eta\to\infty$. If in addition $r\geq3\sqrt{3+2\omega}$ then
the Ricci scalar remains finite at $\eta\to0$ and the evolution
becomes non-singular in the Jordan frame\footnote{Similar behaviour is
found in the axion frame of the low--energy string effective
action~\cite{CEW97}.}. For $\lambda>0$ but $r<\sqrt{3+2\omega}$ the
universe has zero spatial volume as $\eta\to0$ and expands out of the
singularity. For $\lambda<0$ there are no non-singular solutions and
for $r>\sqrt{3+2\omega}$ there is an upper bound on the maximum size
attained by the universe and it undergoes recollapse after a finite
proper time.

Note that in all these solutions there is no cosmological inflation in
the Einstein frame in the sense that $\ddot{\tilde{a}}/
\tilde{a}=h'<0$ at all times.
This must be the case since massless scalar fields do not violate
the dominant energy condition. However,  the solutions for $\eta<0$ do
share many of the useful properties of inflation~\cite{pbb}. In
particular the comoving Hubble length, $1/h=2\eta$, decreases as
$\eta\to0$. This in principle could allow one to explain the homogeneity
of the universe on scales above the Hubble length by causal physics and
this is the basis of the pre--big bang scenario.

In terms of the low--energy string effective action the pre--big bang
epoch~\cite{pbb} is simply a collapsing universe in the Einstein
frame. It is anticipated that in string theory the curvature
singularity which is the usual endpoint of collapse in general
relativity should be avoided due to the large--scale/small--scale
T-duality which implies a minimal length scale. If this can smoothly
connect the collapsing pre--big bang branch to the post--big bang
expanding universe, then string theory could perhaps explain the
large--scale structure of our present universe.

\section{Linear perturbations}
\label{perts}

\setcounter{equation}{0}

\def\theequation{\thesection.\arabic{equation}}

In this Section we consider linear perturbations with 
comoving wavenumber $k$ about the
homogeneous FRW solutions presented in the previous Section. 
Working in the uniform curvature gauge, we can write the linearised
equations of motion as 
\begin{eqnarray}
\delta\phi'' + 2h\delta\phi' + k^2\delta\phi
 &=& 2\lambda e^{2\lambda\phi}
  \left[ \sigma'\delta\sigma' + \lambda\sigma'^2\delta\phi \right]  \ ,\\
\delta\sigma'' + 2h\delta\sigma' + k^2\delta\sigma
 &=& -2\lambda \left[ \phi'\delta\sigma' + \sigma'\delta\phi' \right] \ ,\\
\delta \beta'' + 2h\delta \beta' +k^2\delta \beta &=& 0 \ .
\end{eqnarray}

The curvature perturbation is given from the constraint equation
\begin{equation}
\label{zetacon}
\zeta = {1\over12h} \left( \phi'\delta\phi + \beta' \delta \beta +
e^{2\lambda\phi}\sigma'\delta\sigma \right) \ ,
\end{equation}
where $\zeta$ gives the curvature perturbation on uniform energy density
hypersurfaces as $k\eta\to0$~\cite{CEW97,MFB92}. Technical details
regarding the choice of gauge and definition of metric perturbations
were presented in~\cite{CEW97}.

Although we have a non-trivial coupling between the axion and dilaton
fields, we can in fact integrate the equations of motion to yield
analytic solutions for the linear perturbations, even when the
background fields evolve as given by Eqs.~(\ref{axia}--\ref{axibeta}). To
do so we must exploit the symmetry of the Lagrange density under
S-duality.

We construct explicitly S-duality invariant perturbations
\begin{eqnarray}
\label{defu}
u &\equiv& {\tilde{a}\over2\lambda^2h} {\rm tr} ( JN'J\delta N ) \ ,\\
 &=& {\tilde{a} \over h}
 \left( \phi' \delta\phi + e^{2\lambda\phi}\sigma' \delta\sigma \right)
 \ ,\\
\label{defv}
v &\equiv& {\tilde{a}\over2\lambda^2h} {\rm tr} ( -JNJN'J\delta N ) \ ,\\
 &=& {e^{\lambda\phi} \tilde{a} \over h}
 \left( \phi' \delta\sigma - \sigma' \delta\phi \right) \ .
\end{eqnarray}

The equations of motion for the fields then reduce to
\begin{eqnarray}
u'' + \left[ k^2 + {1 \over 4\eta^2} \right] u &=& 0 \ ,\\
v'' + \left[ k^2 + {1- 4\lambda^2r^2 \over 4\eta^2} \right] v &=& 0 \ .
\end{eqnarray}

The general solution for the axion-dilaton perturbations is thus given
by
\begin{eqnarray}
u &=& |k\eta|^{1/2} \left[ u_+ H_0^{(1)}(|k\eta|) + u_-
H_0^{(2)}(|k\eta|) \right] \ ,\\
v &=& |k\eta|^{1/2} \left[ v_+ H_{|\lambda r|}^{(1)}(|k\eta|) + v_-
H_{|\lambda r|}^{(2)}(|k\eta|) \right] \ ,
\end{eqnarray}
where $H_\nu^{(i)}$ are Hankel functions of the first or second kind
of order $\nu$.  This displays the classic behaviour expected of
perturbations in a FRW model. For small scales ``within the horizon''
(i.e., $|k\eta|\gg1$) the solutions oscillate, while for $|k\eta|\ll1$
they cease oscillating as $|k\eta|\to0$.

Perturbations in the modulus field remain decoupled from the axion
and dilaton fields and their equation of motion can be written as
\be
w'' + \left[ k^2 + {1\over 4\eta^2} \right] w = 0 \ ,
\ee
where
\be
w \equiv {\tilde{a}\over h}  \beta' \delta\beta \ .
\ee
The general solution for linear perturbations in the modulus field is
thus given by
\be
w = |k\eta|^{1/2} \left[ w_+ H_0^{(1)}(|k\eta|) + w_-
H_0^{(2)}(|k\eta|) \right] \ .
\ee
The curvature perturbation $\zeta$ is then given by the constraint
equation~(\ref{zetacon}) as
\be
\label{zeta}
\zeta = {u+w \over 12\tilde{a}} \ 
\ee
and is manifestly S-duality invariant. Because $v$ does not contribute 
to $\zeta$ we can consider it as an isocurvature perturbation. 

For solutions with $\eta>0$ we have a priori no means by which to
determine the constants $u_\pm$, $v_\pm$ and $w_\pm$. All modes start
outside the horizon at the singularity as $\eta\to0$, so there is no
causal mechanism that can establish a particular set of initial
conditions. However for solutions with $\eta<0$, all modes start within
the horizon at early times as $\eta\to-\infty$ and so we may reasonably
assume that at sufficiently early times, $|k\eta|\gg1$, they are in
their vacuum state. 

Note that as $|\eta|\to\infty$ the background solutions approach the
$r_\pm=\pm r$ dilaton--moduli--vacuum solution where $\sigma'\to0$ and
hence, using Eqs.~(\ref{vacuumsol}) and~(\ref{vacuumphisol}),
\begin{eqnarray}
u \to \pm 2r \tilde{a} \delta\phi \ ,\\
v \to \pm 2r e^{\lambda\phi} \tilde{a} \delta\sigma \ .
\end{eqnarray}

Thus $u$ and $v$ reduce to the dilaton and axion perturbations about the
dilaton-vacuum solutions. Using the Minkowski vacuum state to normalise these
in the low curvature regime as $k\eta\to-\infty$ yields~\cite{CEW97,BD}
\begin{eqnarray}
u_+ &=& \pm 2r\, e^{i\pi/4} {\sqrt{\pi} \over 2\sqrt{k}} \
, \qquad u_-=0 \ ,\\
v_+ &=& \pm 2r\, e^{i(2|r\lambda|+1)\pi/4} 
 {\sqrt{\pi} \over 2\sqrt{k}} \ 
, \qquad v_-=0 \ .
\end{eqnarray}

At late times as $\eta\to0$ we again find $\sigma'\to0$ and the
background solutions approach the vacuum solutions with $r_\mp=\mp r$. Thus
$u$ and $v$ again reduce to the dilaton and axion perturbations and at
late times for the $\eta<0$ solutions we can write down the power
spectra for $-k\eta\ll1$:
\begin{eqnarray}
\label{Pphi}
{\cal P}_{\delta\phi} \to {2\over\pi^3} \tilde{H}^2 (-k\eta)^3
[\ln(-k\eta)]^2 \ ,\\
\label{Psigma}
{\cal P}_{\delta\sigma} \to \left( {C(|\lambda r|)\over2\pi}
\right)^2 \left( {k \over e^{\lambda\phi} \tilde{a}} \right)^2
(-k\eta)^{1-2|\lambda r|} \ ,
\end{eqnarray}
where $\tilde{H}\equiv h/\tilde{a}$ is the Hubble expansion rate in the
Einstein frame. The numerical coefficient
\begin{equation}
C(x) \equiv {2^x\Gamma(x) \over 2^{3/2}\Gamma(3/2)} \ 
\end{equation}
approaches unity as $x\to3/2$.
Similarly we find
\be
{\cal P}_{\delta\beta} \to {2\over\pi^3} \tilde{H}^2 (-k\eta)^3
[\ln(-k\eta)]^2 
\ee
for the modulus field. 

In the pre--big bang scenario the linear perturbations are normalised in
the low curvature, weakly--coupled regime as $\eta\to-\infty$. The
pre--big bang era then establishes perturbations on all scales outside
the horizon as $\eta\to0$. Assuming there is a rapid, smooth transition
to the post--big bang epoch this perturbation spectrum then determines
the large--scale structure of the $\eta>0$ universe.

The spectral index of the perturbations is conventionally denoted by $n_x$
where
\begin{equation}
n_x-1 \equiv {d{\cal P}_{\delta x} \over d\ln k} \ ,
\end{equation}
and a scale-invariant Harrison-Zel'dovich spectrum
corresponds to  $n_x=1$.
Thus for the dilaton and axion perturbations as $\eta\to0$ we find from
Eqs.~(\ref{Pphi}) and~(\ref{Psigma}) that 
\begin{eqnarray}
n_\phi &=& 4 \ ,\\
n_\sigma &=& 4 - 2|\lambda r|
 = 4 - 2\left|\lambda\sqrt{3-s^2}\right| \ ,\\
n_\beta &=& 4 \ .
\end{eqnarray}

Note that unlike the dilaton and modulus spectrum, 
the axion spectrum can be consistent with scale invariance. We employ these
results in the following section to derive the perturbation spectra 
generated in a particular string inspired model. 

\section{A string inspired example}
\label{GSsect}

\setcounter{equation}{0}

\def\theequation{\thesection.\arabic{equation}}

Ten--dimensional supergravity theories represent the low energy limit of
string theories. The NS--NS bosonic sector for theories of this type is
\be
\label{ten}
S=\int d^{10}x \sqrt{-g} e^{-\hat{\Phi}} \left[ R + 
\left( \nabla \hat{\Phi} \right) 
-\frac{1}{12} H^2 \right]  \ , 
\ee
where $H_{\mu \nu \lambda} = \partial_{[\mu}B_{\nu \lambda]}$ 
is the rank--three antisymmetric 
tensor field strength and $\hat{\Phi}$ 
is the dilaton field. This can be dimensionally reduced to
four dimensions 
by compactifying on an isotropic six--dimensional torus 
\cite{witten,gs}, where 
the ten--dimensional metric is given by 
\be
\label{tenmetric}
g_{MN}=  \left( \begin{array}{cc} g_{\mu\nu}(x) & 0 \\ 
0 & {\rm exp} [\beta (x)/\sqrt{3}] \delta_{mn} \end{array} \right)   \ ,
\ee
$x$ represents the four--dimensional coordinates,
and $\mu, \nu, \ldots$ $(m, n, \ldots )$  are the four--
(six--) dimensional indices. It is assumed that  the components of the
antisymmetric tensor field are  given by $B_{\mu\nu}=B_{\mu\nu}(x)$ and
$B_{mn} = \alpha(x) \epsilon_{mn}$\cite{witten}. 
It is further assumed that $\hat{\Phi} =\hat{\Phi} (x)$.
 
The effective four--dimensional action 
can then be written as
\be
\label{10to4}
S=\int d^4 x \sqrt{-g} e^{-\phi} \left[ R + \left( \nabla \phi \right)^2 
-\frac{1}{2} 
e^{2\phi} \left( \nabla \sigma \right)^2 
-\frac{1}{2} \left( \nabla \beta \right)^2
-\frac{1}{2} e^{-2 \beta/\sqrt{3}} 
\left( \nabla \alpha\right)^2 \right]  \ ,
\ee
where $\phi \equiv \hat{\Phi} -\sqrt{3} \beta$ 
is the shifted dilaton field and the antisymmetric tensor 
field has been expressed in terms of its dual, the pseudo--scalar 
axion field $\sigma$, i.e., $H^{\mu\nu\lambda} =
e^{\phi} \epsilon^{\mu\nu\lambda\rho} \nabla_{\rho} \sigma$, 
where $\epsilon^{\mu\nu\lambda\rho}$ is the antisymmetric 
four--form satisfying $\nabla_{\kappa} \epsilon^{\mu\nu\lambda\rho}
=0$.  

By performing the conformal transformation Eq.~(\ref{conformal}) to the
Einstein frame, it can be seen that the fields $\phi$ and $\sigma$
parameterise an SL(2,R)/U(1) coset as in the 
action Eq.~(\ref{einsteinactionone}) with $\lambda=1$:
\be
\label{q}
S = \int d^4x \sqrt{-\tilde{g}} \left[ \tilde{R} 
-\frac{1}{2} \left( \tilde{\nabla} \phi \right)^2 -
\frac{1}{2} e^{2 \phi} \left( \tilde{\nabla} \sigma \right)^2 
-\frac{1}{2} \left( \tilde{\nabla} \beta \right)^2 -
\frac{1}{2} e^{-2 \beta/\sqrt{3}} \left( \tilde{\nabla} \alpha \right)^2
 \right]  \ .
\ee
However, the reflection symmetry of the 
modulus field $\beta$ is extended. 
The moduli fields $\alpha$ and $\beta$ together parameterise another
SL(2,R)/U(1) coset with an effective coupling $\lambda = - 1/\sqrt{3}$. 
The action Eq.~(\ref{q}) may therefore be written as 
\be
\label{action*}
S = \int d^4x \sqrt{-\tilde{g}} \left[ \tilde{R} 
+ {1\over4} 
{\rm Tr} \left( \tilde{\nabla} N \tilde{\nabla} N^{-1} \right)
+ {3\over4}
{\rm Tr} \left( \tilde{\nabla} P \tilde{\nabla} P^{-1} \right) \right]
\ ,
\ee
where $N$ is given by  Eq.~(\ref{S}) with $\lambda=1$, and 
\be
P = 
\left( \begin{array}{cc} e^{-\beta/\sqrt{3}} &  {1\over\sqrt{3}}\alpha
e^{-\beta/\sqrt{3}} \\
 {1\over\sqrt{3}} \alpha e^{-\beta/\sqrt{3}} &
 e^{\beta/\sqrt{3}} +  {1\over3}\alpha^2 e^{-\beta/\sqrt{3}}
\end{array} \right) \ .
\ee
The Lagrange density in the Einstein frame is 
therefore invariant under two independent
SL(2,R) transformations 
\be
\label{theta}
\tilde{g}_{\mu\nu} \to \tilde{g}_{\mu\nu}\ , \quad
N\to \Theta N \Theta^T \ ,\quad
P\to \theta P \theta^T \ ,
\ee
where $\Theta$ and $\theta$ are two independent SL(2,R) matrices.

When $\alpha$ is frozen, the action Eq.~(\ref{action*}) reduces to
Eq.~(\ref{einsteinactionone}) with $\lambda =1$. Thus the general
cosmological solution for $\alpha'=0$ is given by
Eqs.~(\ref{axia})--(\ref{axibeta}). 
The SL(2,R) transformation $P\to\theta P\theta^T$ then yields the general
moduli field solutions with $\alpha'\neq0$,
\bea
\label{newbeta}
e^{\beta} =2^{\sqrt{3}} e^{\beta_*} \left[ \left| \frac{\eta}{\eta_*} 
\right|^{-s/\sqrt{3}} + \left| \frac{\eta}{\eta_*} \right|^{s/\sqrt{3}} 
\right]^{-\sqrt{3}}  \ , \\
\label{newc}
\alpha=\alpha_*  \pm \sqrt{3} e^{\beta_* /\sqrt{3}} \left[ 
\frac{|\eta /\eta_* |^{-s/\sqrt{3}} - |\eta /\eta_*|^{s/\sqrt{3}}}{|
\eta/\eta_*|^{-s/\sqrt{3}} + | \eta /\eta_* |^{s /\sqrt{3}}} \right] \ . 
\eea
The transformation mixes these two fields in a non--trivial fashion  and
results in a time--dependent $\alpha$, but it leaves the other fields,
and hence the metric in the string frame as well as the Einstein frame,
invariant. This follows since the  combined role of $\{ \alpha, \beta
\}$ is effectively that of the single modulus field $\beta$ present in
action Eq.~(\ref{actionone}). 

To study linear perturbations about these homogeneous solutions we
should pick quantities invariant under the duality transformations in
Eq.~(\ref{theta}). Thus we have $u$ and $v$ as defined in
Eqs.~(\ref{defu}) and~(\ref{defv}), with $\lambda=1$, and
\begin{eqnarray}
\label{neww}
w &\equiv& {3\tilde{a}\over2h} {\rm tr} ( JP'J\delta P ) \ ,\\
 &=& {\tilde{a}\over h}
 \left( \beta' \delta\beta + e^{-2\beta/\sqrt{3}}\alpha' \delta\alpha
 \right) \ ,\\
x &\equiv& {3\tilde{a}\over2h} {\rm tr} ( -  JPJP'J\delta P ) \ ,\\
 &=& {e^{-\beta/\sqrt{3}} \tilde{a} \over h}
 \left( \beta' \delta\alpha - \alpha' \delta\beta \right) \ .
\end{eqnarray}
The curvature perturbation $\zeta$ is still given by Eq.~(\ref{zeta}), 
but $w$ is now given by Eq.~(\ref{neww}).
 
By analogy with the axion and dilaton perturbations in
Section \ref{perts}, we can immediately 
write down the spectral indices of the
perturbations in this model. We find
\bea
n_\phi=4 \ , \quad n_\sigma = 4 - 2r \ , \\
n_\beta=4 \ , \quad n_\alpha = 4 - {2\over\sqrt{3}}s \ ,
\eea
where $r^2+s^2=3$. 
It is worth noting that the constraint between $r$ and $s$ has the 
effect of linking the indices for the 
two axions $\sigma$ and $\alpha$. 
As $r$ ranges from $0$ to $\sqrt{3}$, we find $n_\sigma$ ranges from 
$4$ to $0.5$ and $n_\alpha$ from $2$ to $4$. 

\section{Discussion}

\setcounter{equation}{0}

\def\theequation{\thesection.\arabic{equation}}

In this paper we have investigated the evolution of dilaton, moduli
and axion (curvature and isocurvature) perturbations in a general
class of models with an arbitrary coupling, $\lambda$, between the
dilaton field and the axion field.  Such models possess an SL(2,R)
symmetry between the dilaton and axion fields making it possible to
solve the full homogeneous equations of motion for the system.  Armed
with these solutions we solved the linear perturbation equations for
the fields, making use of the fact that we could write down unique
S-duality invariant combinations of perturbed fields.  In a pre--big
bang type scenario the dilaton and modulus spectra, and thus from
Eq.~(\ref{zeta}) the curvature perturbation, is strongly tilted
towards small scales, leaving an almost perfectly homogeneous universe
on large scales. While this is consistent with the assumption of a
homogeneous background, it does not provide a spectrum of
perturbations on large scales capable of seeding the density
fluctuations observed today. However the perturbation spectrum of the
axion field is scale invariant when $|\lambda|r = 3/2$, where $r$ is
an integration constant satisfying $0 \leq r \leq \sqrt{3}$.  The
bounds on the possible values of the spectral index are therefore 
\be
\label{minimumtilt}
4-2\sqrt{3}|\lambda| \leq n_\sigma \leq 4. 
\ee
Hence it is possible to obtain $n_\sigma \le 1$ for $|\lambda| \ge 
\sqrt{3}/2$. 

String theory offers many possibilities in terms of the low--energy
four--dimensional theory. Axion type fields can arise 
from the compactification of the
antisymmetric three--form field strength on the internal space. We
have investigated the consequences of an isotropic 
compactification where the extra axion, $\alpha$, 
couples to the modulus field such that the full action has an 
SL(2,R)$\times$SL(2,R) 
symmetry. This enables us to calculate the evolution of the 
perturbations in the more
complicated system. Whereas the
dilaton and moduli spectra produced in the pre--big bang scenario 
are too steep, both
axion fields could have less steep spectra. Their
spectral indices are linked through the Friedmann constraint
equation. In particular as $r$ ranges from $0$ to
$\sqrt{3}$, $n_\sigma$ ranges from $4$ to $0.5$ and $n_{\alpha}$ from 
$2$ to $4$. Such a result indicates that only the $\sigma$
field can possess a scale-invariant spectrum in this model.

The perturbation spectra are generated in the pre--big bang phase. 
Perturbations inherited by the post--big bang phase will depend on the 
mechanism by which the transition between the two phases 
proceeds. However, despite this
caveat, it is important that in principle it is possible to generate
observationally interesting power spectra in this scenario, and it
demonstrates the importance of axion fields in the pre--big bang
picture.

\end{document}